 \documentclass[aps,preprint,superscriptaddress,showpacs]{revtex4-1}

\usepackage{pifont}

\usepackage{graphicx}
\usepackage{indentfirst,latexsym,bm}
\usepackage{txfonts}
\preprint{APS/123-QED}
\begin{document}


\title{Indirect-direct hybrid-drive work-dominated hotspot ignition for inertial confinement fusion}



\author{X. T. He}
\affiliation{Institute of Applied Physics and Computational Mathematics, P.O.Box 8009, Beijing 100094, China}
\affiliation{Center for Applied Physics and Technology, HEDPS, and IFSA Collaborative Innovation Center of MoE, Peking University, Beijing 100871, China}
\affiliation{Institute of Fusion Theory and Simulation, Zhejiang University, Hangzhou 310027, China}


\author{Z. F. Fan}
\affiliation{Institute of Applied Physics and Computational Mathematics, P.O.Box 8009, Beijing 100094, China}

\author{J. W. Li}
\affiliation{Institute of Applied Physics and Computational Mathematics, P.O.Box 8009, Beijing 100094, China}
\affiliation{Center for Applied Physics and Technology, HEDPS, and IFSA Collaborative Innovation Center of MoE, Peking University, Beijing 100871, China}

\author{J. Liu}
\affiliation{Institute of Applied Physics and Computational Mathematics, P.O.Box 8009, Beijing 100094, China}
\affiliation{Center for Applied Physics and Technology, HEDPS, and IFSA Collaborative Innovation Center of MoE, Peking University, Beijing 100871, China}

\author{K. Lan}
\affiliation{Institute of Applied Physics and Computational Mathematics, P.O.Box 8009, Beijing 100094, China}
\affiliation{Center for Applied Physics and Technology, HEDPS, and IFSA Collaborative Innovation Center of MoE, Peking University, Beijing 100871, China}

\author{J. F. Wu}
\affiliation{Institute of Applied Physics and Computational Mathematics, P.O.Box 8009, Beijing 100094, China}

\author{L. F. Wang}
\affiliation{Institute of Applied Physics and Computational Mathematics, P.O.Box 8009, Beijing 100094, China}
\affiliation{Center for Applied Physics and Technology, HEDPS, and IFSA Collaborative Innovation Center of MoE, Peking University, Beijing 100871, China}

\author{W. H. Ye}
\affiliation{Institute of Applied Physics and Computational Mathematics, P.O.Box 8009, Beijing 100094, China}
\affiliation{Center for Applied Physics and Technology, HEDPS, and IFSA Collaborative Innovation Center of MoE, Peking University, Beijing 100871, China}


\date{\today}

\begin{abstract}
An indirect-direct hybrid-drive work-dominated hotspot ignition scheme for inertial confinement fusion is proposed: a layered fuel capsule inside a spherical hohlraum with an octahedral symmetry is compressed first by indirect-drive soft-x rays (radiation) and then by direct-drive lasers in last pulse duration. In this scheme, an enhanced shock and a follow-up compression wave for ignition with pressure far greater than the radiation ablation pressure are driven by the direct-drive lasers, and provide large $pdV$ work to the hotspot to perform the work-dominated ignition. The numerical simulations show that the enhanced shock stops the reflections of the indirect-drive shock at the main fuel-hotspot interface, and therefore significantly suppresses the hydrodynamic instabilities and asymmetry. Based on the indirect-drive implosion dynamics the hotspot is further compressed and heated by the enhanced shock and follow-up compression wave, resulting in the work-dominated hotspot ignition and burn with a maximal implosion velocity of $\sim400 km/s$ and a lower convergence ratio of $\sim25$. The fusion yield of $15 MJ$ using total laser energy of $1.32 MJ$ is achieved.
\end{abstract}


\maketitle


%
In a central hotspot-ignition scheme of inertial confinement fusion (ICF)\cite{Nuckolls72,Atzeni04}, a layered capsule with a frozen deuterium-tritium (DT) main fuel shell is imploded, the central hotspot is compressed and heated to reach ignition condition. A thermonuclear burn wave from the ignited hotspot propagates outward towards the high compressed main fuel, resulting in the self-sustaining burn and the fusion energy gain. Two main implosion schemes, direct drive (DD)\cite{Bodner98} and indirect drive (ID)\cite{Lindl95}, have been proposed. In the DD scheme, lasers are directly focused on the capsule ablator surface. In the ID scheme, the capsule is placed inside a hohlraum with a high-Z shell and lasers irradiate on an inner wall of the hohlraum to generate radiation (soft x-rays) that ablates and heats the capsule ablator surface, where a high temperature and pressure plasma is formed, resulting in implosion dynamics to highly compress the main fuel and heat the hotspot performing ignition. So far, the hotspot is designed in isobaric ignition, for which the implosion shock running inside the hotspot experiences multiple reflections back and forth from the hotspot center to an interface between the hotspot and high compressed main fuel (below, called the hotspot interface) till stagnation that the inward velocity at the hotspot interface vanishes. It results in the fact that at the hotspot interface each reflection exerts a force from a light fluid (the hotspot) to an inward-accelerating heavy fluid (the high-compressed density main fuel), and then the heavy fluid is decelerated, where $\nabla p\cdot\nabla \rho<0$ with $p$ pressure and $\rho$ density, and consequently the hydrodynamic instabilities\cite{Taylor50} and mix happen since the actual implosion shock is not an ideal spherical symmetry. The shock reflections also amplify the asymmetry of the hotspot interface since there always exists some spatial roughness at the interface, resulting in the hotspot distortion. In the recent full ID experiments on the National Ignition Facility (NIF), the layered capsule is put inside a cylindrical hohlraum that may cause a radiation flux asymmetry at the CH ablator surface of the capsule and exacerbate the hotspot distortion, in addition, the isobaric-ignition hotspot driven by the low ablation pressure most in $\sim 100 Mbar$ and the low implosion velocity most in $\sim300 km/s$\cite{Haan11,Lindl14,Hurricane14} may undergo serious hydrodynamic instabilities and mix. In the low adiabat ($\alpha= P/P_F\sim1.45$) experiments\cite{Lindl14}, where $P$ and $P_F$ are the main-fuel pressure and Fermi pressure at stagnation respectively, the implosion dynamics caused serious distortion of the hotspot and main-fuel layer, and ablator materials mixed into the hotspot, resulting in the failure of the hotspot isobaric ignition. In the high adiabat ($\alpha\sim2.8$) experiments\cite{Hurricane14,Dittrich14}, although the hydrodynamic instabilities at the radiation ablation front (RAF) of the CH ablator were improved and the hotspot performed the isobaric ignition, no burn appeared and the hotspot distortion and mix were clearly observed. It has exhausted the existing NIF energy of $\sim2 MJ$. In addition, in the high adiabat experiments the main fuel only arrived at the lower areal density ($<1.0g/cm^2$), which is difficult to perform burn with gain.

In this report, we proposed an ID-DD hybrid-drive work-dominated ignition scheme to solve the above issues. The capsule, which is put in a spherical hohlraum with an octahedral symmetry to ensure radiation symmetry, is first compressed by the successively intensive ID shocks that adiabatically compress the main fuel to high density and are combined into a merged shock (MS) in the vicinity of the inner surface of the main fuel layer. The ID MS runs into the hotspot that is first compressed and heated, and rebounds from the hotspot center to the hotspot interface. On the other hand, an enhanced shock (ES) with the pressure greater than that of the MS is launched for ignition by DD lasers\cite{He13,*Fan13}, and rapidly arrives at the hotspot interface and collides with the MS that is stopped before its first reflecting there. Therefore, the hydrodynamic instabilities and asymmetry caused by the MS at the hotspot interface are suppressed. The ES and a follow-up compression wave through the interface provide large $pdV$ work to the hotspot that is further compressed and heated, resulting in the work-dominated hotspot ignition soon after the ES running in the hotspot first reflects at the hotspot interface. During implosion compression, the rate of the total internal energy $E$ in the hotspot varies from $dE/dt<0$ to $dE/dt>0$. The hotspot ignition condition can be calculated by using $dE/dt =0$. Assume that the temperature $T$ and density $\rho$ in the hotspot are homogeneous, we obtain the equation for the ignition condition ($\alpha$-particle self-heat beginning)\cite{Atzeni04,Lindl95} for a DT system under spherical symmetry geometry in the form
\begin{eqnarray}
g(T,\xi ){(\rho {r_D})^2} + {A_W}\Gamma Tu(\rho {r_D}) = {A_S}\frac{{{T^{7/2}}}}{{\ln \Lambda }},\label{Eq1}
\end{eqnarray}
it indicates a relationship between the areal density $\rho r_D(g/cm^2 )$ and temperature $T (keV)$, where $r_D$ is the hotspot radius. In Eq.(\ref{Eq1})
\begin{eqnarray}
g(T,\xi ) = {A_Q}n_T^2\xi (1 - \varphi ){\left\langle {\sigma \,v} \right\rangle _{DT}}{\varepsilon _0}{f_\alpha } - {A_B}n_T^2{(1 + \xi )^2}\sqrt T, \label{Eq2}
\end{eqnarray}
is the net increase rate ($g>0$) of $\alpha$-particle energy deposition (the first term in the right side of Eq.(\ref{Eq2})) taking off electron bremsstrahlung emission energy (the second term). In Eq.(\ref{Eq2}), ${f_\alpha } = 1 - \frac{1}{{4{\theta _\alpha }}} + \frac{1}{{160\theta _\alpha ^3}}$ with ${\theta _\alpha } = \rho {r_D}/\rho {\ell _\alpha}$ is a fraction of the $\alpha$-particle energy deposited into the hotspot\cite{Krokhin73} and $\rho {\ell _\alpha } = \frac{{0.025{T^{5/4}}}}{{1 + 0.0082{T^{5/4}}}}$ is the $\alpha$-particle range at densities $\rho=(10-100)g/cc$\cite{Atzeni04,Hurricane14}, and a fraction of bremsstrashlung x-ray energy absorbed in the hotspot is neglected. $g(T,\xi)=0$ is a boundary between the $\alpha$-particle energy deposition rate in thermonuclear reaction and the electron energy emission loss rate, where $\xi=n_D/n_T$ is a ratio of the deuterium particle number $n_D$ to the tritium $n_T$. The multi-deuterium system is very important for ICF because the ion temperature during burning process can far exceed the deuterium-deuterium ignition threshold, in which the tritium is generated and thus the initial assembled tritium is saved. In the ignition discussion, we take the burn-up fraction $\phi=0$ and $\xi=1$, and from $g=0$ we can get the minimal ignition temperature $T\sim4.3keV$. In Eq.(\ref{Eq1}), $A_W\Gamma T$ comes from the power $pu$ provided by the ES and compression wave through the hotspot interface, where the pressure $p=\Gamma \rho T$, the velocity $u$ at the hotspot interface and the isothermal-sound velocity $C_T = \sqrt{\Gamma T}$ are both in 10km/s, $A_sT^{7/2}/ln\Lambda$ is from the electron thermal conduction rate, $<\sigma v>_{DT}$ is the DT thermonuclear reaction rate in $10^{-24}cm^3/s$, $ln\Lambda$ is the Coulomb logarithm, and $\varepsilon_0$ is the $\alpha$-particle energy in $MeV$. The coefficients $A_Q =1.6\times10^6$, $A_B=5.0\times10^5$, $A_W=34.8$, $A_S =2.7\times10^8$. Eq.(\ref{Eq1}) is a quadric in form since ${\theta _\alpha } = \rho {r_D}/\rho {\ell _\alpha }\sim 1$ in $f_{\alpha}(\theta_{\alpha})$. We have the general solution from in the form
\begin{eqnarray}
\rho {r_D} = \frac{{{A_w}\Gamma Tu}}{{2g}}(\sqrt {1 + h}  - 1) \equiv {f_1}(T,u), \label{Eq3}
\end{eqnarray}
where $h = \frac{{4g{A_s}{T^{7/2}}/\ell n\Lambda }}{{A_w^2{\Gamma ^2}{T^2}{u^2}}}$.

\begin{figure}[h]
\includegraphics[width=0.5\textwidth]{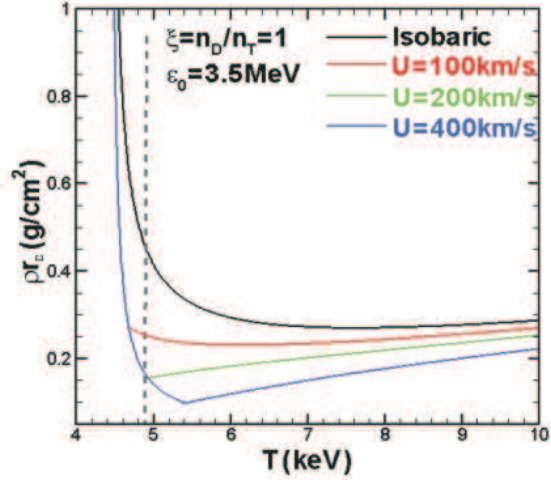}%
\caption{\label{Fig1} (Color online) Hot-spot ignition conditions: the areal density $\rho r_D$ versus temperature T for $\xi=1$ and   particle energy $\varepsilon_0= 3.5 MeV$. The curves show the isobaric ignition (u=0, black solid line), and the nonisobaric ignition with velocities u (km/s)=100 (red), 200(green), and 400(blue).}
\end{figure}

Let $u=0$, the isobaric ignition threshold is
\begin{eqnarray}
\rho {r_D} = \sqrt {\frac{{{A_s}{T^{7/2}}/\ell n\Lambda }}{g}}  \equiv {f_2}(T). \label{Eq4}
\end{eqnarray}
Let $u$ larger enough to make $h<<1$, we have the work-dominated ignition threshold
\begin{eqnarray}
\rho {r_D} = \frac{{{A_s}{T^{7/2}}/\ell n\Lambda }}{{{A_w}\Gamma Tu}} \equiv {f_3}(T,u). \label{Eq5}
\end{eqnarray}
Eq.(\ref{Eq5}) has an additional relation of $A_w \Gamma Tu = {2g}(T,\xi  = 1)$, substituting it into Eq.(\ref{Eq1}) we can obtain Eq.(\ref{Eq5}) again. Therefore, from it or $h<<1$ we could estimate the interface velocity $u=u(T)$ demanded for the work-dominated ignition, then combining Fig.\ref{Fig1} we could get the ignition threshold. We usually call Eq.(\ref{Eq3}) nonisobaric ignition except for the case of the isobaric ignition, while the work-dominated ignition is nonisobaric without the MS reflections at the hotspot interface.

It is seen from Fig.\ref{Fig1} that for the same temperature the isobaric ignition demands higher $\rho r_D$ to compare with the work-dominated ignition. For example, taking $T\sim4.9 keV$, the work-dominated ignition ($u=400 km/s$) demands only $\rho r_D \sim0.15 g/cm^2$ while the isobaric ($u=0$) demands $\rho r_D\sim0.45 g/cm^2$, as shown by the vertical dashed-line in Fig.\ref{Fig1}. It is as well shown that for lower velocity, such as $u=100 km/s$, the ignition is close to isobaric. In fact, the inward velocity $u$ at the hotspot interface is close to the implosion velocity $V_{im}$, where the main fuel is in high compressed density, therefore, $V_{im}\sim400 km/s$ is necessary to promote the hotspot temperature and perform the work-dominated hotspot ignition and main fuel burn.

Such high implosion velocity $V_{im}$ is difficult to be provided only by the ID ablation pressure. In our hybrid scheme, this implosion velocity for ignition is driven by the DD lasers that are steadily imposed on a critical surface in the corona plasma formed by the ID ablated ablator inside the SH. A supersonic electron thermal wave with an electron ablation front (EAF) propagates from the critical surface inward towards the RAF, and decays across the corona plasma where the wave energy is deposited. Once the supersonic velocity slows down to an electron isothermal sonic $C_s=(\Gamma_eT_e)^{1/2}$ at time $t=t_s$, an electron ablation shock and a follow-up steady electron compression wave with high pressure at the EAF are launched, where $T_e$ is the electron temperature and $\Gamma_e$ is the electron pressure coefficient. The electron compression wave continually drives the corona plasma inward toward the RAF like a snowplow. Meanwhile, the ablator is continually ablated by the radiation temperature at the RAF as well. Therefore, a high density plateau between the EAF and RAF is formed due to the density pile-up of the corona plasma and ablated ablator plasma. We now discuss how the corona plasma is driven by the electron compression wave using the snowplow model. Assume that the areal plasma density (mass per unit area) at time $t$ accumulated in front of the EAF (here and below, the EAF means the EAF of the electron compression wave, unless otherwise specified) is $\mu (\tau ) = {\mu _0}(0) + {\mu _1}(\tau )$ with $\tau=t-t_s$, where ${\mu _0}(0) = \int\limits_{{R_{RA}}(0)}^{{R_{EA}}(0)} \rho dR$ is the areal density between the radius $R_{EA}(\tau=0)$ of the EAF and radius $R_{RA}(\tau=0)$ of the RAF, ${\mu _1}(\tau ) = \int\limits_0^\tau  {\dot m d\tau }$ is the areal density of the ID ablated mass of the ablator at the RAF, where the mass ablation rate $\dot{m}\sim T^3$\cite{Olson11}. The motion of the corona plasma pushed by the EAF can be described by the snowplow equation:
\begin{eqnarray}
\frac{d}{{d\tau }}(\mu {R^2}\frac{{dR}}{{d\tau }}) =  - {R^2}[{P_{EA}}(\tau ) - P(\tau )],\label{Eq6}
\end{eqnarray}
where $P_{EA}$ is a high pressure ahead of the electron compression wave and $P=\Gamma \rho T$ is the corona plasma pressure. Taking initial conditions: $R(\tau=0)=R_{EA}(0)$ and $dR/d\tau(\tau =0)\approx0$, the radius $R_{EA}$ of the EAF is obtained from the Eq.(\ref{Eq6}), i.e., ${R_{EA}}(\tau ) = {R_{EA}}(0) - \int\limits_0^\tau  {\frac{{d{\tau _1}}}{{\mu ({\tau _1}){R^2}({\tau _1})}}\{ } \int\limits_{{0}}^{{\tau_1}} {d{\tau _2}} {R^2}({\tau _2})[{P_{EA}}({\tau _2}) - P({\tau _2})]\}$. Consider that the velocity at the RAF is approximately the imploding velocity, we can write the plateau width between radius $R_{EA}$ of the $EAF$ and radius $R_{RA}$ of the $RAF$ in the form
\begin{eqnarray}
\Delta {R_{ER}}(\tau ) = \Delta {R_{ER}}(0)
&& + \int\limits_0^\tau  {d{\tau _1}\{ {V_{im}} - } \frac{1}{{{\mu _1}({\tau _1}){R^2}({\tau _1})}}\int\limits_0^{\tau_1}  {d{\tau _2}} {R^2}({\tau _2})[{P_{EA}}({\tau _2}) - P({\tau _2})]\},\label{Eq7}
\end{eqnarray}
where $\Delta R_{ER}(\tau )=R_{EA}(\tau )-R_{RA}(\tau)$ and $\Delta R_{ER}(0)=R_{EA}(0)-R_{RA}(0)$. In early stages, $P_{EA}$ driven by the steady DD lasers is close to a constant, at the EAF, $P_{EA}/P= \Gamma_e T_e/\Gamma T>>1$, where the electron temperature $T_e$ is far greater than the radiation temperature $T$, and $V_{im} \sim T^{1/2}$\cite{Olson11} at the RAF is less than the driving velocity by $P_{EA}$, therefore, $V_{im}<\frac{1}{{{\mu _1}({\tau _1})}}\int\limits_0^\tau  {d{\tau _2}} {P_{EA}}({\tau _2})\sim C_S=(\Gamma_e T_e)^{1/2}$. As a result, the snowplow pushes the density plateau width to be contracted, i.e., $\Delta R_{ER}(\tau) <\Delta R_{ER}(0)$, and thus the averaged plasma density $<\rho>\sim \mu_0/\Delta R_{ER}(\tau)$ goes up within the density plateau. With the lapse of time, later, $P=\Gamma \rho T>>P_{EA}$ due to $T$ and $\rho$ rising while $T_e$ decreasing within the $\Delta R_{ER}(\tau)$, and thus $\Delta R_{ER}(\tau)> \Delta R_{ER}(0)$. However, noticing that ${\mu_1}(\tau )\sim\int\limits_0^\tau  {{T^3}d\tau }$, $V_{im}\sim T^{1/2}$ and $P/\mu \sim T^{-2}$, as a result, the averaged plasma density $<\rho>\sim \mu/\Delta R_{ER}(\tau)$ increases rapidly within $\Delta R_{ER}(\tau)$, where an extremely high pressure $P=\Gamma \rho T$ far greater than the ID ablation pressure $P_a (\sim 100 Mbar)$ is produced. Thus, the analytical discussion showed that the DD lasers may drive an extremely high pressure plateau between the RAF and EAF, and it may drive the ES toward the implosion capsule with the implosion velocity greater than $400 km/s$ to perform the work-dominated hotspot ignition.

\begin{figure}[h]
\includegraphics[width=0.2\textwidth]{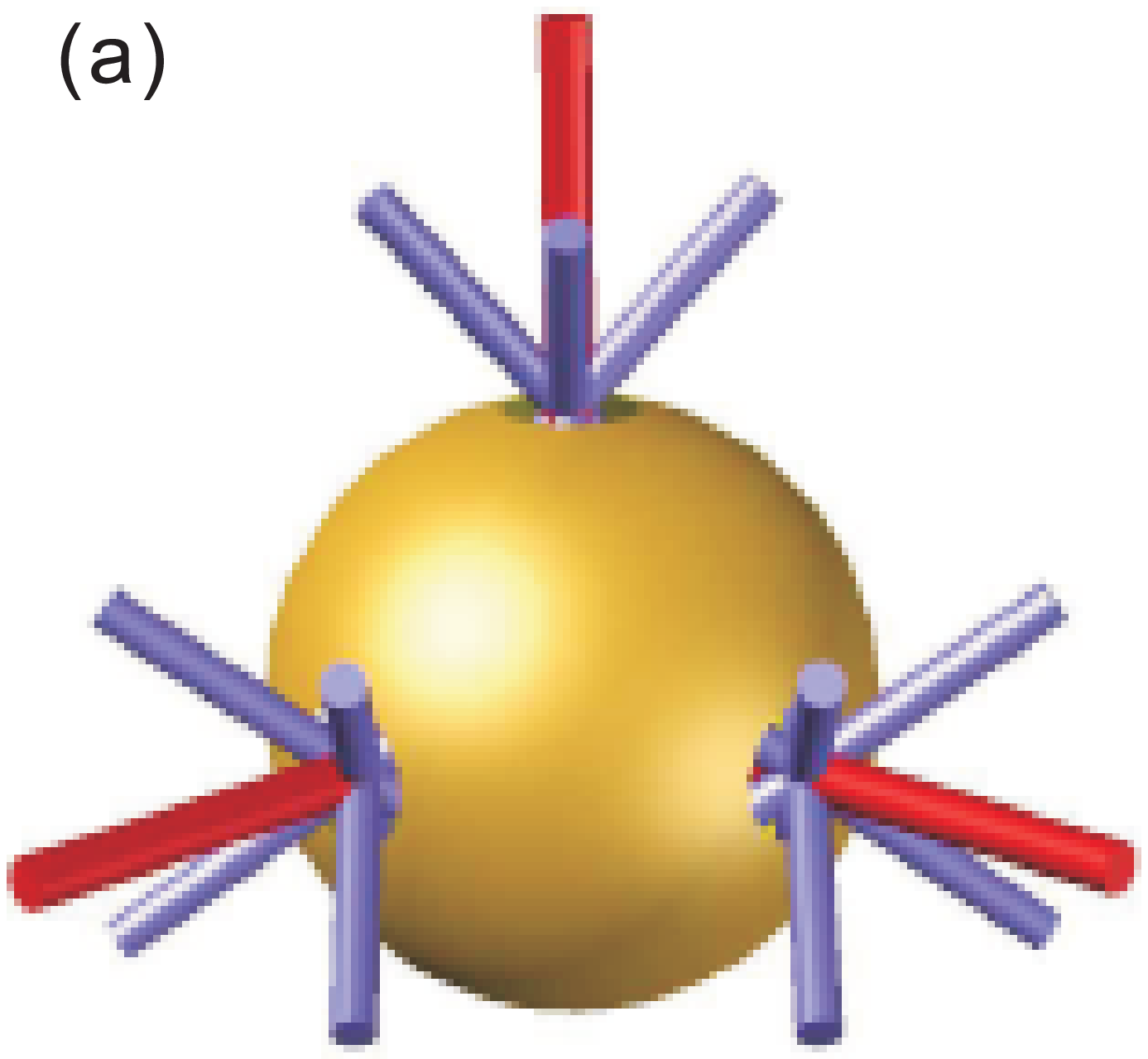}%
\includegraphics[width=0.25\textwidth]{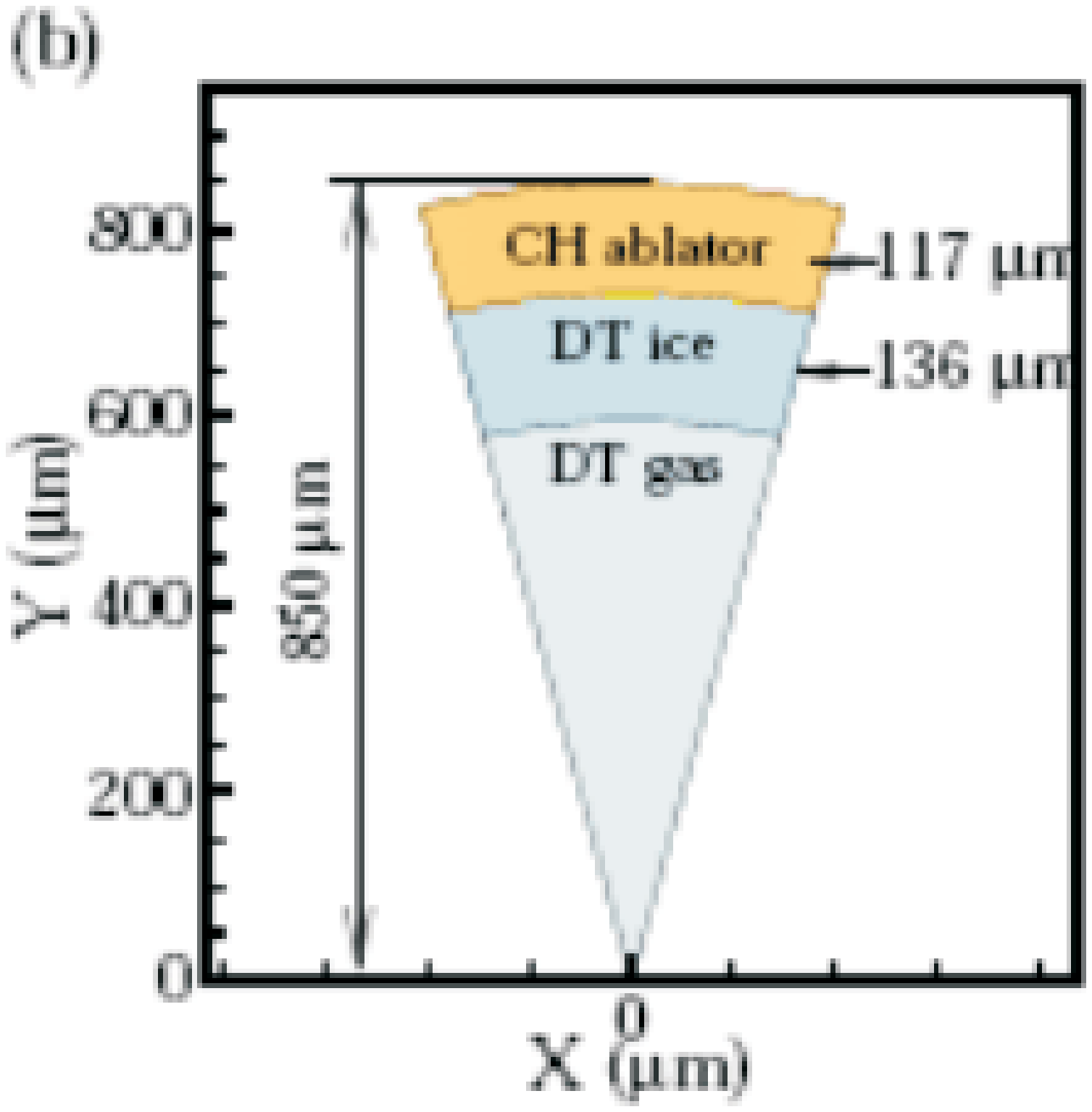}
\includegraphics[width=0.3\textwidth]{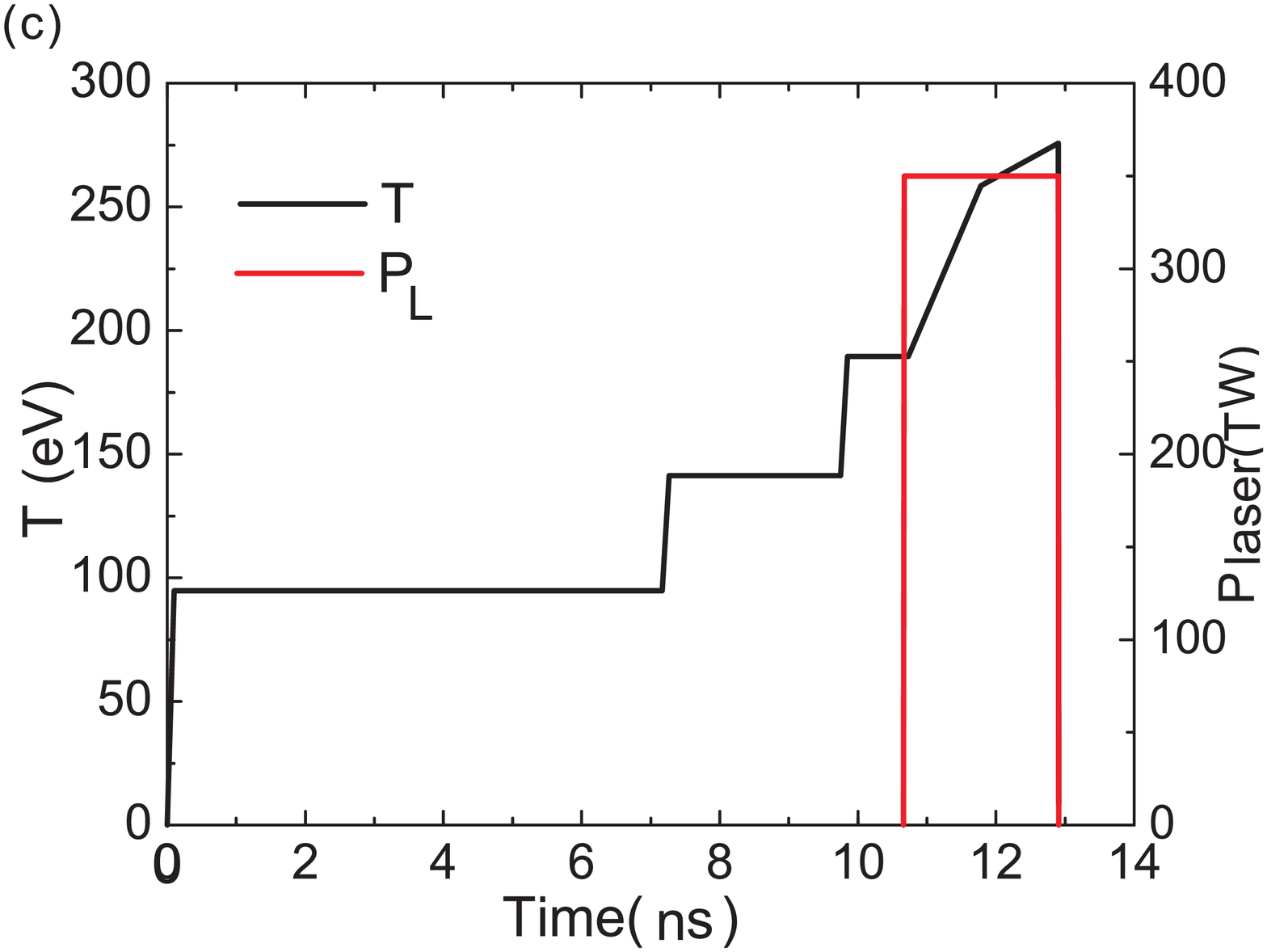}
\caption{\label{Fig2}(Color online) Configurations of the SH and layered capsule, and the evolution of the ID temperature and DD laser power. (a) The schematic plot of the SH with six laser entrance holes (LEHs) and octahedral symmetry; (b) the capsule structure: CH ablator (the outer radius of $0.85 mm$), DT ice layer, DT gas cavity; (c) the radiation temperature T(eV, black) with four steps and the steady DD laser power $P_L$ of $350 TW$ (red) in the pulse duration of last $2.2 ns$.}
\end{figure}

In order to conduct the numerical simulations, we specify the target structure and the source to drive implosion dynamics. The SH filled an electron density of $0.05 n_c$ has six laser entrance holes (LEHs) with an octahedral symmetry (Fig.\ref{Fig2}a)\cite{Lan14,*Lan14-2} and the radii are $4.0 mm$ for the SH and $0.9 mm$ for the LEH. The area ratio of the LEHs to the hohlraum, an important factor characterizing the soft x-ray loss from the LEHs, is $\sim 7\% $ for the SH comparable to the cylindrical hohlraum ($5.75 mm$ in diameter, $9.4 mm$ in length, $3.7 mm$ in LEH's diameter at two ends, and a volume close to the SH) in high-foot NIF experiments\cite{Kline13}. In addition, in the SH, laser beams incident on the inner wall are in a single ring for each LEH, i.e., laser-beam spots on the inner wall are distributed spatially on a ring, therefore no laser beam overlap and cross transfer appear in the SH of the 6 LEHs, and calculation shows that the x-ray flux asymmetry of $\Delta F/F\sim 0.5\%$ is mainly from the spherical harmonics $Y_{44}$ and $Y_{40}$\cite{Lan14,*Lan14-2}. It is of better radiation symmetry and higher coupling of lasers to target compared with the cylindrical hohlraum in which lasers are incident in inner and outer rings to tune the complicated time-dependent radiation flux asymmetry to ensure the hotspot symmetry\cite{Town14}. The profile of the capsule in the SH is made up of a CH ablator with an outer radius of $850 \mu m$, a cryogenic DT main fuel layer with mass of $0.19 mg$, and a cavity filled DT gas with density about $0.3 mg/cc$, as shown in Fig.\ref{Fig2}b. The capsule is imploded and compressed first by the ID temperature that has four-step temporal distribution from $100 eV$ to peak $270 eV$ (Fig.\ref{Fig2}c), which drives four successively intensive shocks, in a total pulse duration of $12.9 ns$ (Fig.\ref{Fig2}c), corresponding to the ID laser energy of $0.55 MJ$\cite{Lan14,*Lan14-2}. Then the capsule is further compressed by the steady DD lasers of an intensity $2.8\times10^{15} W/cm^2$ in last $2.2 ns$ duration with a flat-top power of $P_L=350 TW$ and laser energy of $0.77 MJ$ (Fig.\ref{Fig2}c).

The simulations are first performed in one-dimensional spherical geometry using the radiation-hydrodynamics code LARED-S, which is multi-dimensional, massively parallel and Eulerian mesh based, and 2000 meshes with minimum grid size of $0.05\mu m$. The LARED-S code includes laser ray- tracing, multi-group radiation diffusion (20 groups), radiation hydrodynamics, electron and ion heat conduction, atomic physics, nuclear reaction, $\alpha-$particle transport, and quotidian equation of state (QEOS).

\begin{figure}[h]
\includegraphics[width=0.8\textwidth]{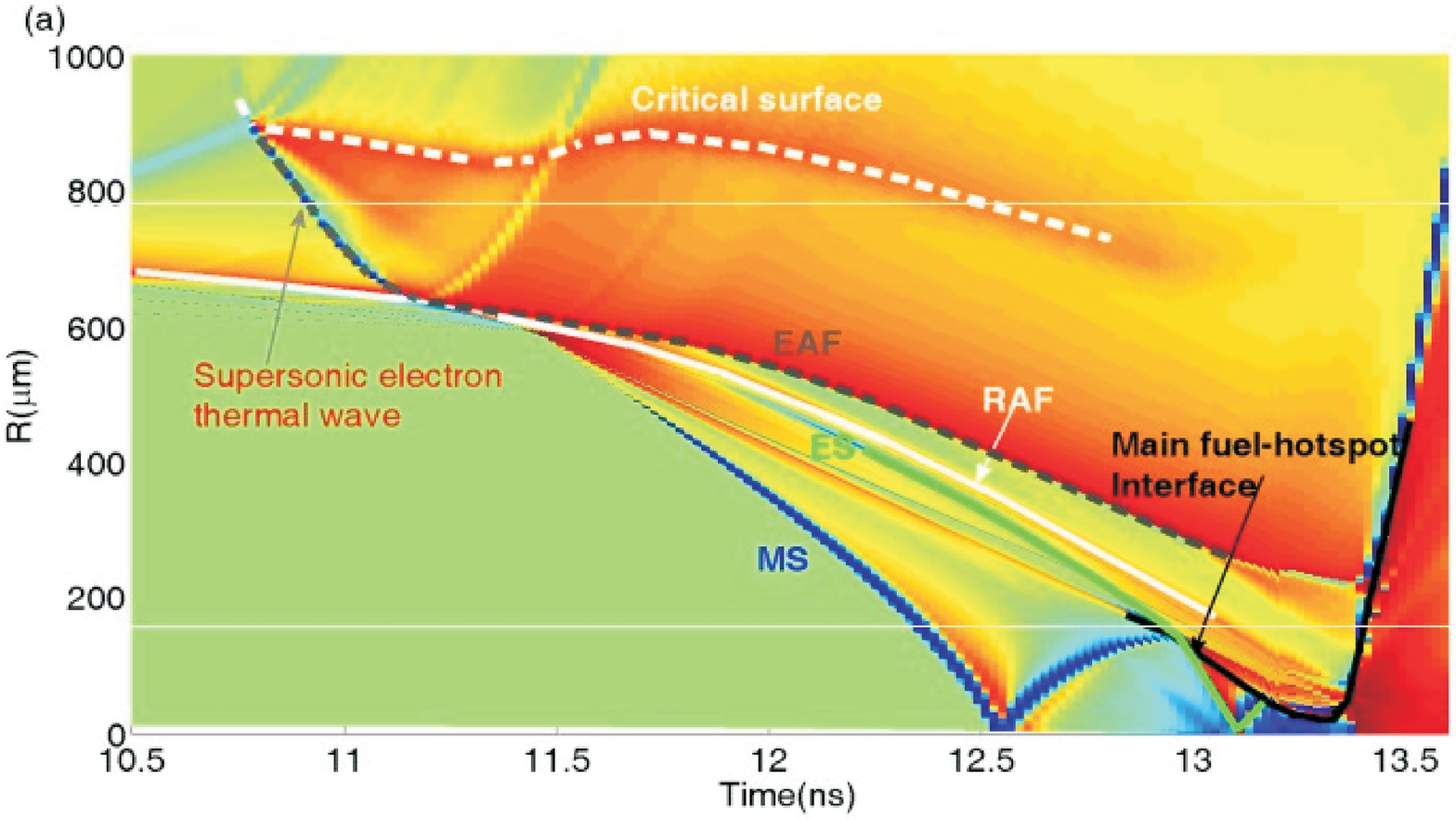}%

\includegraphics[width=0.3\textwidth]{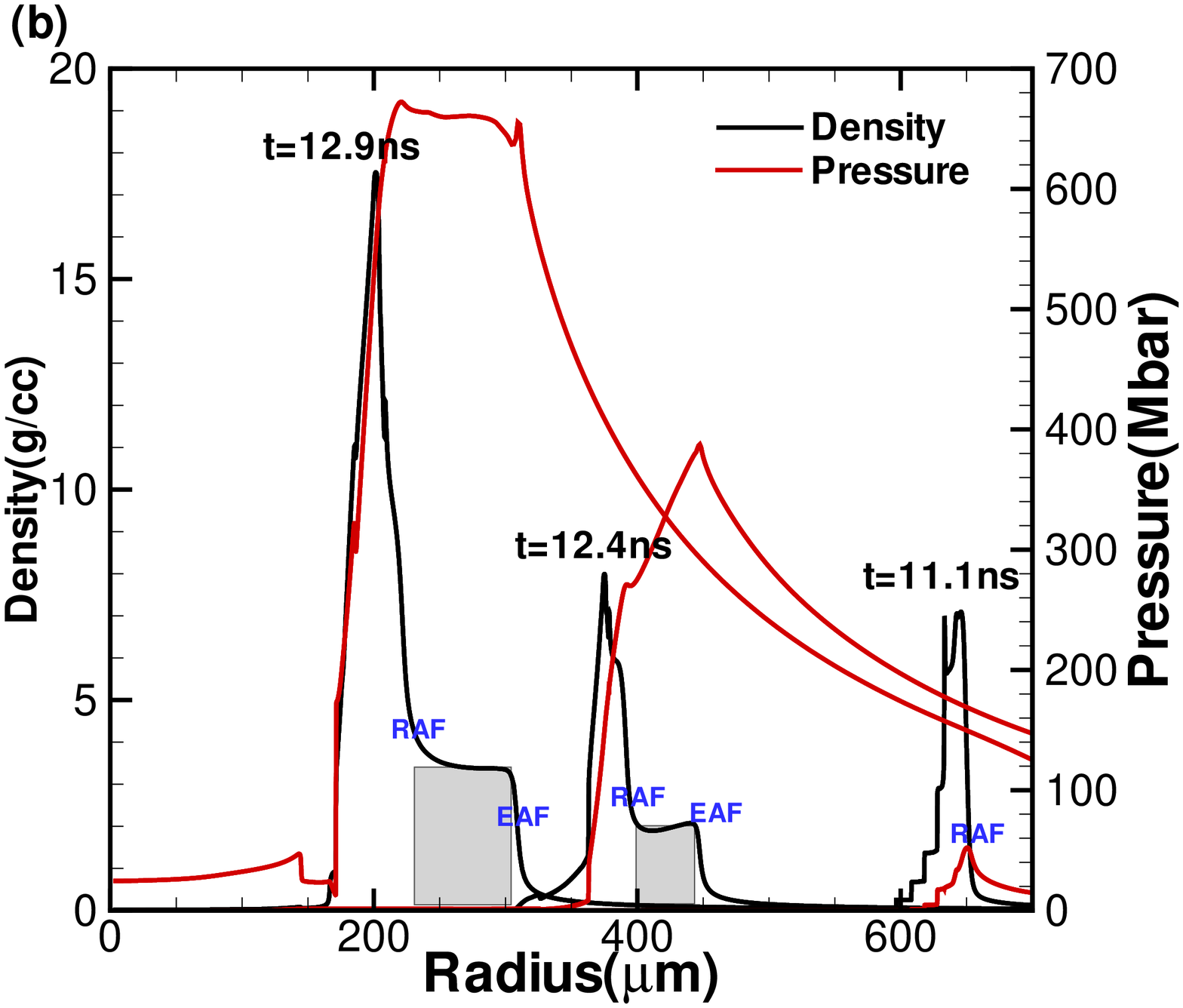}
\includegraphics[width=0.3\textwidth]{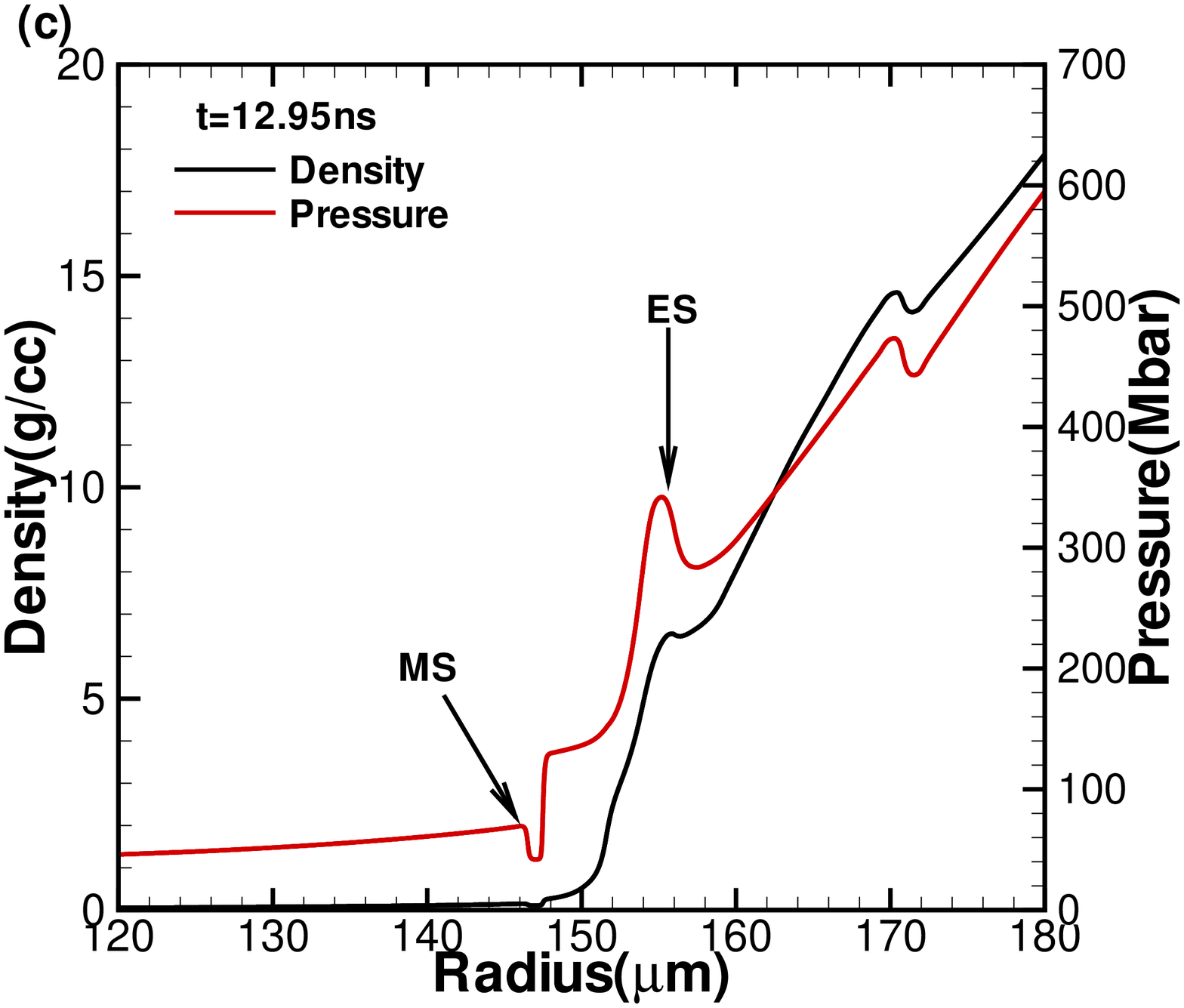}
\includegraphics[width=0.3\textwidth]{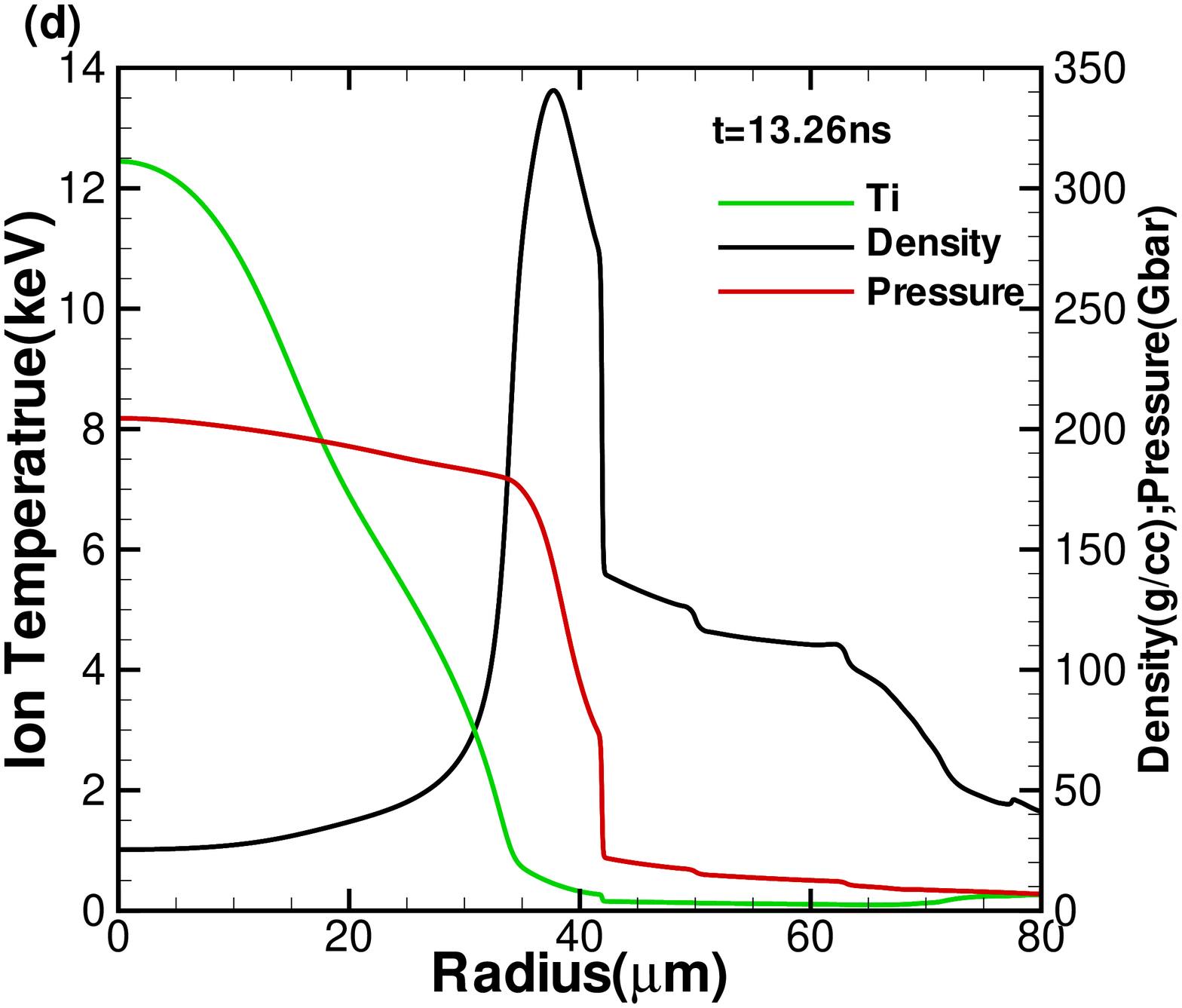}%
\caption{\label{Fig3} (Color online) Implosion physics of the ID-DD hybrid scheme. (a) The radii of interfaces (fronts) in the imploding capsule versus time: the DD laser critical surface(white-dashed line), the electron ablation front EAF (dark-gray-dashed line), the radiation ablation front RAF (white-solid line), the enhanced shock ES (green line), the merged shock MS (blue line), the hotspot boundary (black line); (b) the ID ablation pressure of $50 Mbar$ (red color) at $t\sim 11.1 ns$ before the EAF arriving; the high density and pressure plateaus (gray region) between the EAF and RAF at $t\sim 12.4 ns$ and $12.9 ns$, respectively; (c) the ES and the MS collision occurring at the hotspot boundary, density (black) and pressure (red); (d) ion temperature (green), density (black) and pressure (red) at $t\sim 13.26ns$ (ignition time).} 
\end{figure}

Numerical results of implosion dynamics are shown in Fig.\ref{Fig3}, where the radii of the interfaces (fronts) of the imploding capsule varying with time are indicated. The RAF (white solid line in Fig.\ref{Fig3}a) formed by the ID temperature ablating the CH is rapidly converged during implosion dynamics. The successively intensive four shocks driven by the four-step ID temperature run into the main fuel that is adiabatically compressed, at $t\sim 11.4 ns$ the four shocks and another shock driven by the DD lasers merge into a single MS in the innermost part of the main DT layer which provides the main mass of the hotspot. The convergent MS followed a compression wave runs directly inside the hotspot that is first compressed and heated. During this time, the MS arrives at the hotspot center at $t\sim 12.55 ns$ and consequently rebounds there, and then runs outward toward the hotspot interface at the radius of $\sim 148 \mu m$ at $t\sim 12.95 ns$. On the other hand, at $t\sim 10.7ns$ the steady DD lasers are imposed on the critical surface at the radius of $\sim960 \mu m$ (white dashed-line in Fig.\ref{Fig3}a) in the corona region, which is $\sim 300 \mu m$ away from the RAF. The DD laser energy is strongly absorbed and the electron temperature $T_e$ rises rapidly near the critical surface. The supersonic electron thermal wave is observed, and the electron ablation front EAF propagates in the corona plasma with a maximal velocity of $\sim 850 km/s$ inward toward the RAF. The electron ablation shock and follow-up electron compression wave are launched when the supersonic electron thermal wave slows to a sound speed of $\sim200 km/s$ in the EAF (electron temperature $T_e\sim 1.2 keV$ and ion $T_i\sim 1.0 keV$) at the radius $R_{EA}\sim660 \mu m$ ($15\mu m$ away from the radius $R_{RA}$ of the RAF) at $t\sim 11.1ns$, where the plasma density is $\sim 0.34 g/cc$ and the electron ablation pressure is $P_{EA}\sim240 Mbar$ far greater than the ID ablation pressure of $50 Mbar$ (Fig.\ref{Fig3}b). The electron compression wave with high pressure, like the snowplow as shown by the Eq.(\ref{Eq6}), drives the corona plasma and piles up it into a high density and pressure plateau between the EAF and RAF, where the averaged density of $\sim 2.0 g/cc$ at $t\sim 12.4 ns$ and $\sim 3.3 g/cc$ at $t\sim 12.9 ns$ (gray region in Fig.\ref{Fig3}b) and the peak pressure of $\sim 400 Mbar$ and $\sim 680 Mbar$ respectively (red curves in Fig.\ref{Fig3}b), and the ID temperature is $260\sim270 eV$ there during this period (Fig.\ref{Fig1}c). Such a plateau predicted in analytical discussion above plays a role of a high-pressure piston that drives the ES and steady compression wave. The inward ES runs in the imploding capsule and fast arrives near the hotspot interface at the radius of $148\sim150 \mu m$ at $t\sim12.95 ns$, where the ES is in pressure of $\sim 400 Mbar$ and collides with the MS that is stopped as the MS with pressure of $\sim70 Mbar$ (Fig.\ref{Fig3}c) just arrived there from the hotspot center. Then, the ES directly enters into the hotspot, and it together with the follow-up compression wave further compresses and heats the hotspot, resulting in the ion temperature and pressure rapidly rising. After the ES rebounded from the hotspot center reaches the interface at $t\sim13.18 ns$, soon the hotspot ignition (the heating rate of $\alpha$-particles equals to the cooling rate by electron bremsstrahlung and heat conduction) occurs, that is, the hotspot reaches the mass averaged ion temperature $T_i\sim5.1 keV$, the pressure $\sim 200 Gbar$, the averaged areal density $\rho r_D=0.15 g/cm^2$, the $pdV$ work about $4 kJ$ ($\sim 2/3$ from the ES and compression wave), and total fusion energy of $25 kJ$ is released ($\sim$1/5 deposited in the hotspot) (Fig.\ref{Fig3}d), resulting in the work-dominated ignition in which the hydrodynamic instabilities and hotspot distortion are suppressed. At that time, a convergence ratio (the initial outer radius of capsule to the final radius of the hotspot) is $\sim 25$ that is far less than $\sim 35$ in NIF-NIC experiments\cite{Haan11}. In our scheme, the maximal drive pressure at the RAF is $\sim 700 Mbar$ (far larger than the ID maximal ablation pressure $\sim 100 Mbar$) and the maximal implosion velocity achieves $\sim400 km/s$ which is demanded for the work-dominated ignition as discussed before. In fact, after ignition, the inward velocity u at the hotspot interface is still large enough to be able to further supply the $pdV$ work to the ignited hotspot till stagnation ($u=0$), which, indeed, provides a margin for the hotspot ignition. Until stagnation beginning at $t\sim 13.32 ns$, the isobaric distribution with pressure of $\sim 930 Gbar$ in the ignited hotspot is observed, the system has released the fusion yield of $\sim 200 kJ$ ($\sim 35 KJ$ deposited in the hotspot), the areal density of the main fuel is $1.5 g/cm^2$, $\alpha =1.6$. Finally, the fusion yield of $15 MJ$ under total laser energy of $1.32 MJ$ is achieved.

\begin{figure}[h]
\includegraphics[width=0.3\textwidth]{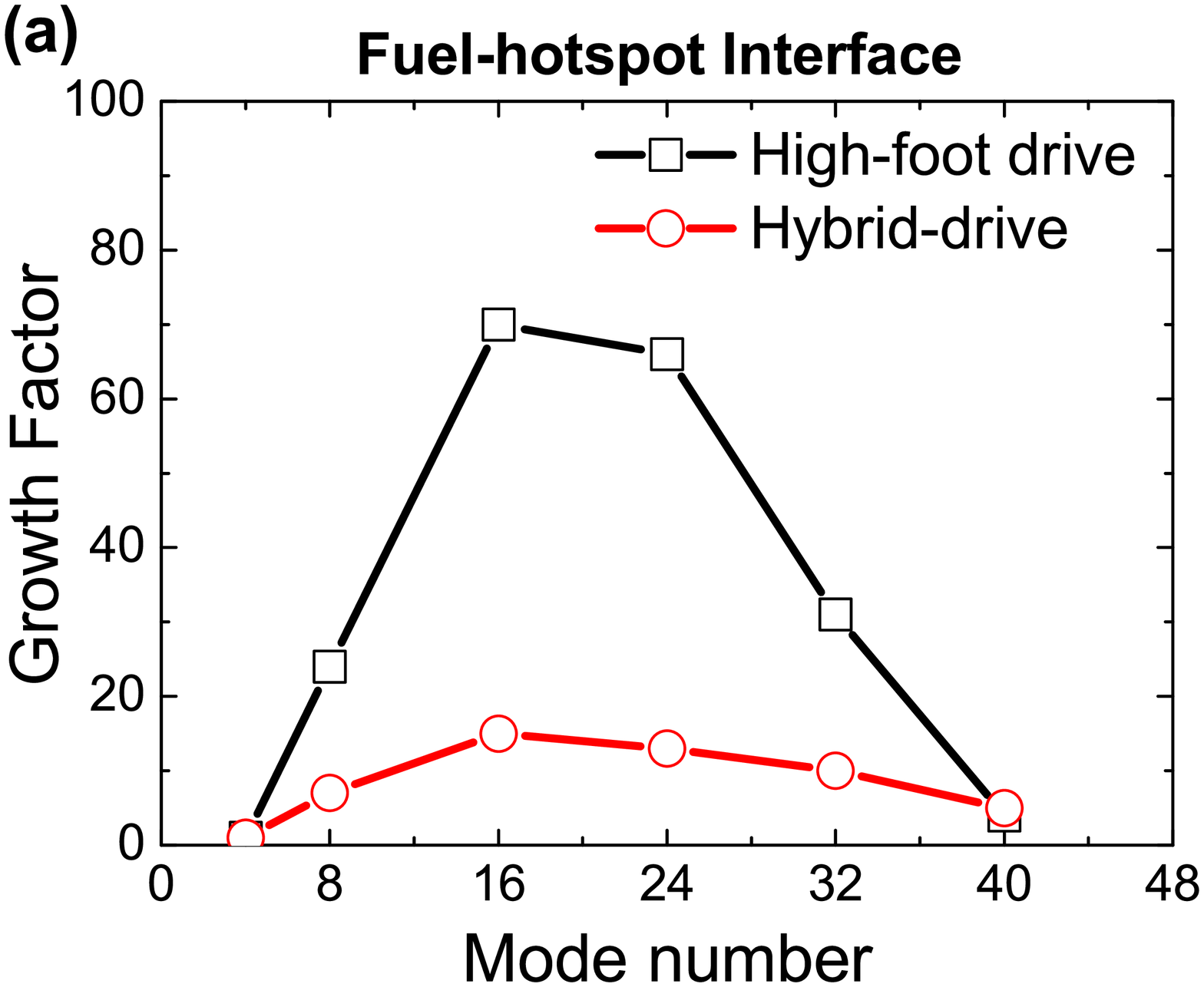}%
\includegraphics[width=0.3\textwidth]{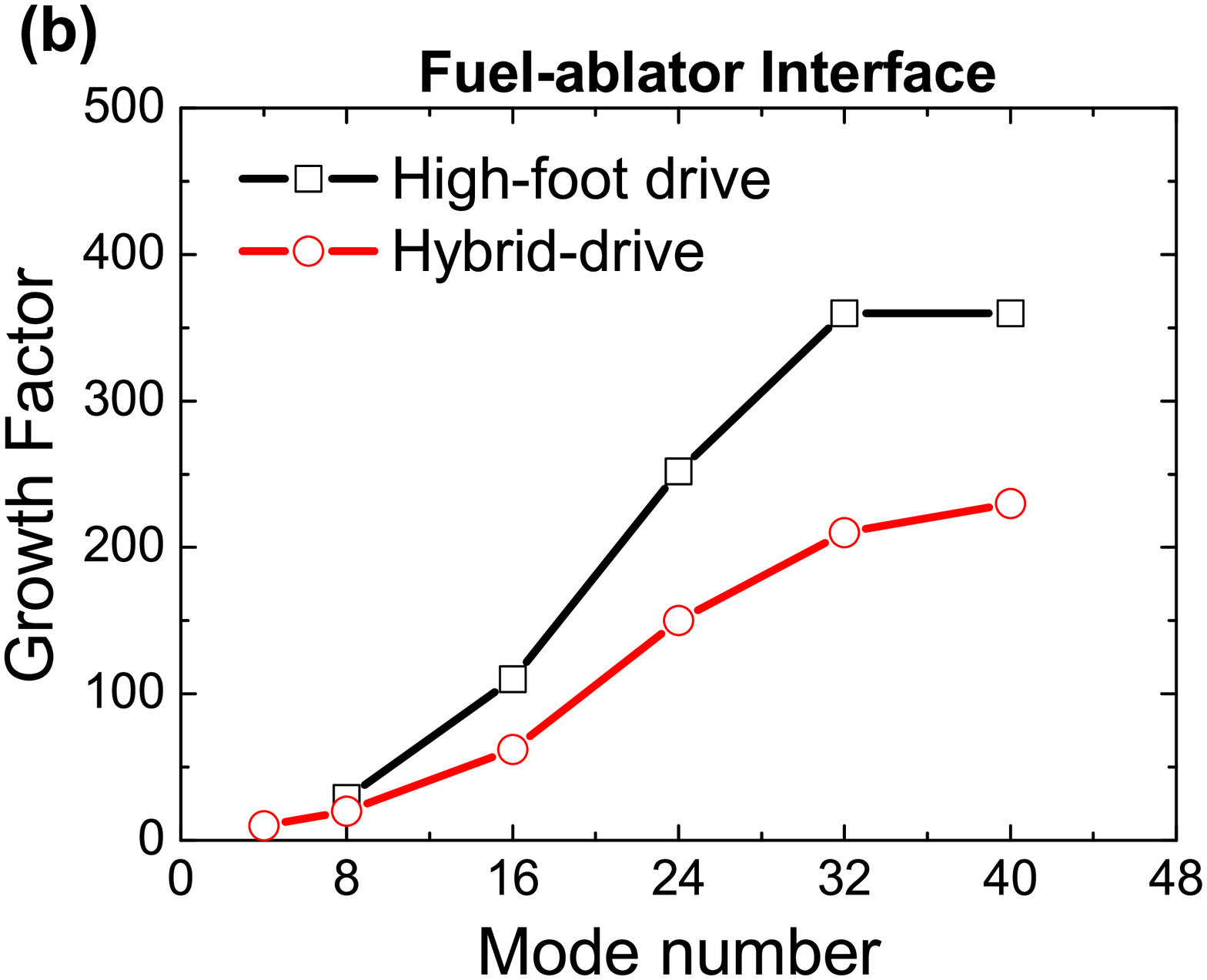}
\includegraphics[width=0.3\textwidth]{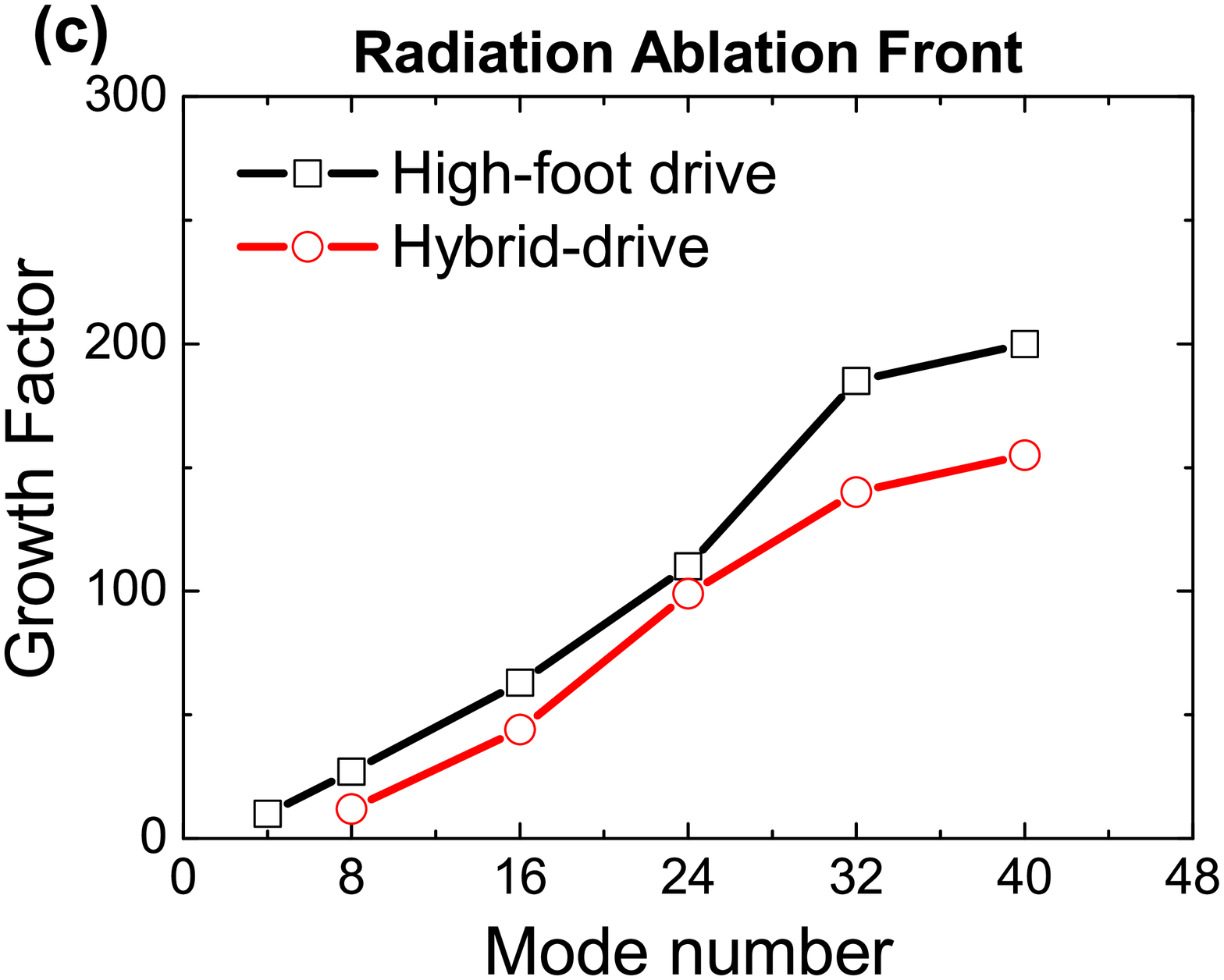}%
\caption{\label{Fig4} (Color online) Comparisons of the GFs of hydrodynamic instabilities (till ignition time) between the hybrid-drive capsule and the ID capsule for perturbations initially seeded at the ablator outer surface, and the 2-dimensional capsule profile at the ignition time. Density contours for a perturbation with a mode $l=16$ and an initial roughness of $350 {\AA}$. (a)-(c) all red curves for present hybrid-drive and all black curves for the high-foot indirect drive in NIF experiments: (a) at the fuel-hotspot interface; (b) at the fuel-ablator interface; (c) at the radiation ablation front (RAF).} 
\end{figure}

We now discuss the two-dimensional LARED-S simulation results. First of all, we investigate the hydrodynamic instabilities at interfaces of the imploding capsule. Define growth factor (GF) as a ratio of the final amplitude to the initial perturbation, Figs.\ref{Fig4}a-c show the GFs at different interfaces varying with the mode $l$ till ignition time. Making a comparison of GFs between the work-dominated hybrid-drive ignition (red curves) and the high-foot hotspot ignition in the NIF experiments (black curves), both are simulated by our LARED-S code. At the RAF the maximal GFs with mode $l=40$ are $\sim150$ for hybrid and $\sim200$ for high-foot\cite{Dittrich14} (Fig.\ref{Fig4}c); at the main fuel-ablator interface the maximal GFs with mode $l=40$ are $\sim 220$ (hybrid) and $\sim 370$ (high foot) in Fig.\ref{Fig4}b, i.e., GFs for hybrid-drive at two interfaces are only $75\%$ and $60\%$ of high foot drive respectively. Most essential is for the GF at the hotspot interface, where the maximal GF is $\sim15$ for hybrid (Fig. 4a) appearing in $l=16$ and is far less than $\sim 70$ of the high-foot ignition in NIF experiments\cite{Dittrich14,Hurricane14}, which only the hotspot ignition occurs without burning due to mix. Our simulations as well manifest that if the ES is absent, soon after the MS first reflects at the hotspot interface, the deceleration is rapidly increasing till stagnation, and the GF magnifies $5\sim10$ times for dominant modes within $\sim 300 ps$, also see\cite{Temporal06}. Thus the work-dominated hybrid-drive ignition scheme can significantly suppress the hydrodynamic instabilities and mix at the hotspot interface by using the ES implosion ignition, and the results show as well that the feedthrough effect from the outer-interfaces is negligible. Therefore, the hotspot is of benefit to performing ignition and burn.
%

\begin{figure}[h]
\includegraphics[width=0.4\textwidth]{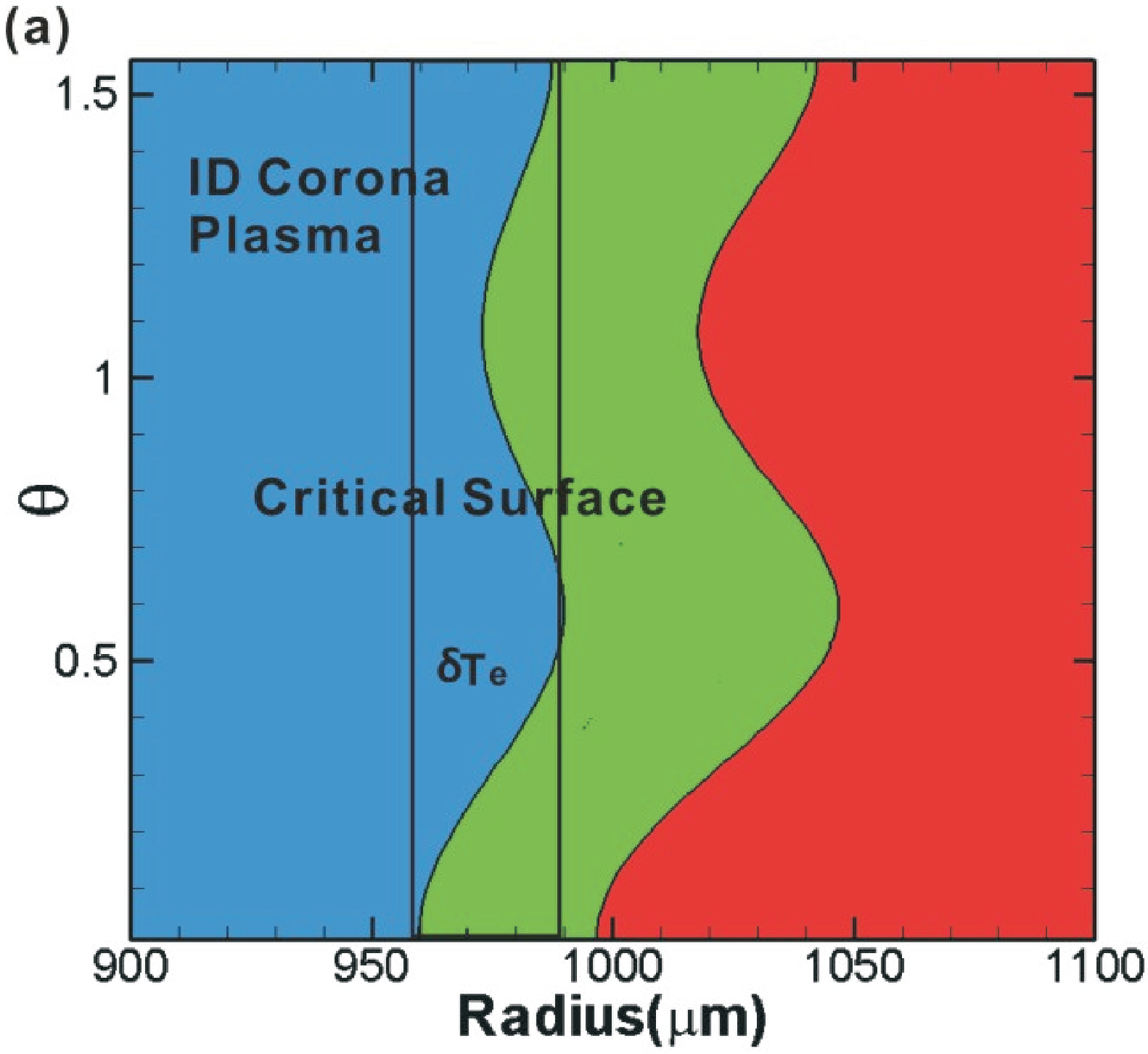}%
\includegraphics[width=0.4\textwidth]{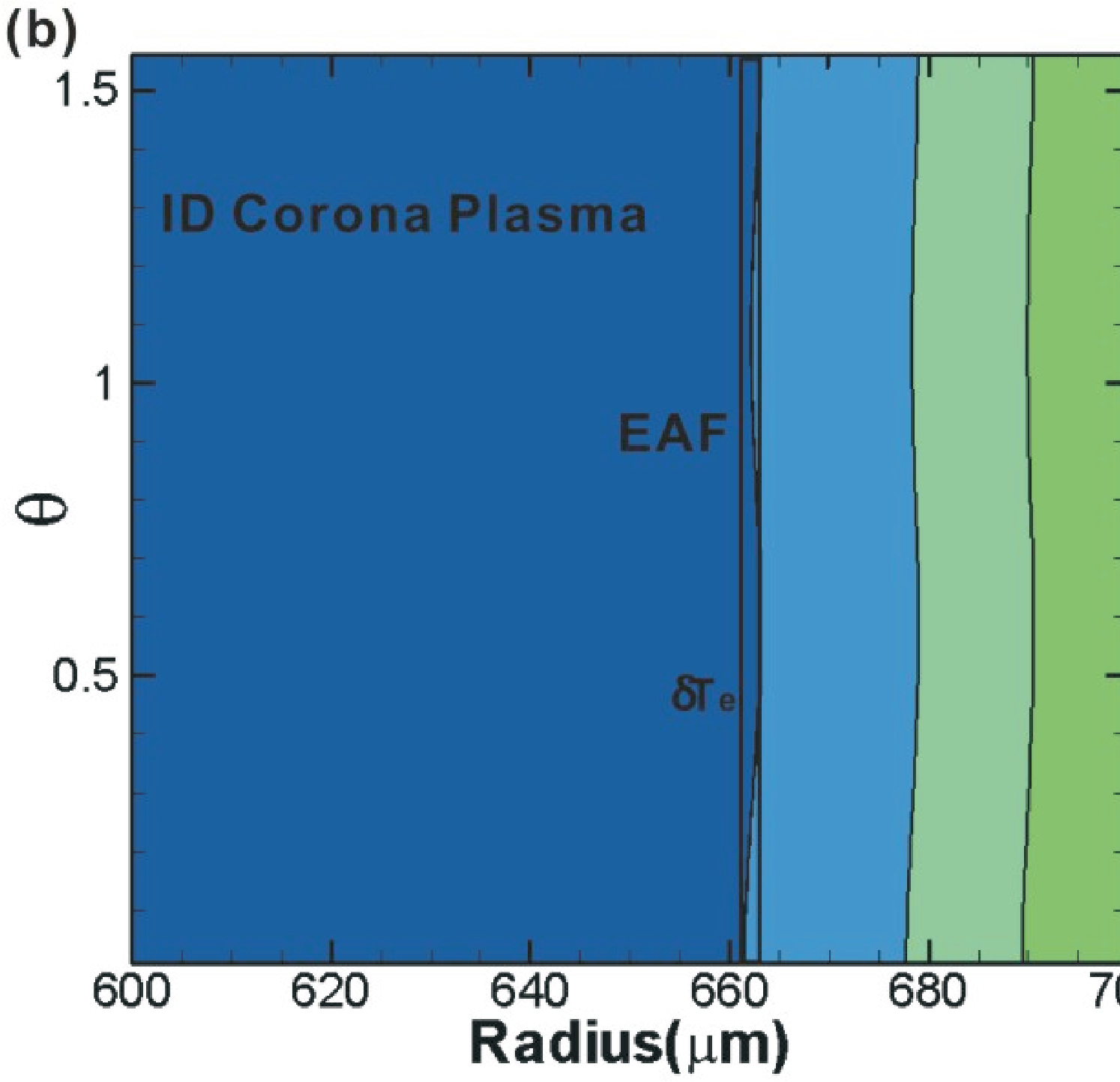}
\caption{\label{Fig5} (Color online) The electron temprature $T_e$ nonuniformity at spherical coordinate of $R-\theta$, which is caused by DD lasers imposed at the critical surface, is suppressed by thermal smoothing of the supersonic electron ablation wave in the corona plasma. (a) Initial temperature perturbation $\delta T_e\sim30 \mu m$ at the critical surface ($R\sim 960\mu m$) at $t\sim10.7 ns$; (b) perturbation amplitude $\delta T_e\sim1.5\mu m$ at the EAF ($R\sim660 \mu m$) at $t\sim 11.1 ns$ when the electron ablation wave slows down to sonic.}
\end{figure}

Hydrodynamic instabilities as well occur at the DD laser critical surface in the corona region, where the non-uniformities from the laser imprinting (smaller scales) and edge overlapping (larger scales) of laser beam bundles would take place. Meanwhile, the supersonic EAF of the electron thermal wave leaves the critical surface and propagates inward towards the RAF at the average velocity of $\sim850 km/s$, and the corona plasma is almost static during propagation. 2-dimensional simulations by LARED-S code show that the initial temperature perturbation $\delta T_e \sim 30 \mu m$ in space (Fig.\ref{Fig5}a), induced by the laser intensity nonuniformity $\sim 10\%$, imposed at the critical surface, is suppressed to $\delta T_e \sim 1.5 \mu m$ from $t\sim 10.7$ to $11.1 ns$ when the electron ablation wave slows down to sonic at the radius of $\sim 660 \mu m$. As a result, the effect of the nonuniformity imposed by the DD laser on the capsule implosion is negligible. Recently, the thermal smoothing in the nonuniform laser-target interaction was also verified by using 2-dimensional fully Fokker-Plank model simulations\cite{Keskinen09} and it was observed as well that this model predicts more smoothing than the hydrodynamic Spitzer mode. In our recent experiments on the $SG-II$ laser facility, significant smoothing effect was observed as well\cite{SGII14}.
%

Laser plasma interaction (LPI) in the ignition target is concerned for two reasons: the laser energy degradation by instabilities that are mainly from the stimulated Raman scattering (SRS), stimulated Brillouin scattering (SBS), two plasmon decay (TPD), and energetic electrons generated by these instabilities, which might preheat the imploding fuel core to degrade its compression. In NIF experiments that all ID laser beams are incident in the cylindrical hohlraum using double rings, the inner ring beams occur in cross and overlap on the optical ways, it results in the laser intensity increase and beam energy transfer, and the SRS is enhanced. Laser beam coupling to the target is $\sim (84\pm3)\%$ in a wide range of laser energy and power\cite{Kline13}, and the energetic electrons with kinetic energy $>100 keV$ are absorbed in about 0.2 mg DT ice (preheat) in an upper bound of $(5 \pm3)J$\cite{Doppner12}, which is $\sim(3.8\pm2.3)\times10^{-6}$ of total laser energy of $1.3MJ$, it results in an adiabat increase of $\delta \alpha\sim3.5\%$ that is acceptable in $\alpha \sim1.5$. Making a comparison, in our hybrid scheme, the layered capsule is inside the SH with the radius of $4 mm$, the ID laser energy coupling to the target is greater than $90\%$ because of only one ring lasers incident, the short length of LPI and no beam overlap and energy cross transfer result in low LPI$\sim6\%$ mainly from SBS measured in OMEGA experiments\cite{Wallace99}. The preheating DT energy by energetic electrons $>100 keV$ is estimated at most $(2.1\pm1.3)J$ for the ID laser energy of $0.55 MJ$, and thus the effect of fuel preheating on the adiabat is negligible. As for the DD lasers, the backscattering experiments on OMEGA regard to the shock ignition\cite{Betti07} for spike beams show that the energy loss of the total DD lasers with intensity of $2.8\times10^{15}W/cm^2$ is $\sim 10\%$ without using phase plates\cite{Theobald12}. In fact, the loss is significantly reduced if using smooth technologies. The further experiment is necessary for the ignition lasers of higher intensity.

In summary, we have proposed the indirect-direct hybrid-drive work-dominated ignition scheme for inertial confinement fusion. The layered fuel capsule inside the octahedral symmetric spherical hohlraum is compressed first by the ID radiation pressure and then by the DD lasers in the last pulse duration. The simulations by LARED-S show that the piston-like high-pressure plateau with pressure far greater than the ID ablation pressure is formed between the EAF and RAF and the ID implosion dynamics is significantly improved. The enhanced shock for ignition is driven by the high-pressure plateau, and the maximal imploding velocity reaches$\sim400 km/s$ demanded for the work-dominated ignition. The ES stops the ID shock reflections at the hotspot interface and suppresses the hydrodynamic instabilities and asymmetry there, thus two-dimensional influence may be significantly improved. The hotspot is further compressed and heated by the ES, resulting in the work-dominated ignition in a lower convergence ratio of 25. Finally, the fusion energy of $15 MJ$ under total laser energy of $1.32 MJ$ is achieved.

We would like to thank C. Y. Zheng, H. B. Cai, Z. S. Dai, L. Hao, and M. Chen at IAPCM, C. K. Li at MIT and J. Fernandez at LANL for discussions. This work was performed under the auspices of the National ICF Program of China, the National Basic Research Program of China (Grant No. 2013CB34100), the National High Technology research and development Program of China(Grant No. 2012AA01A303), and the Nature Science Foundation of China (Grant Nos: 11475032, 11175027, 11391130002, 11274026, and 11205010).


\begin{thebibliography}{10}


 \bibitem{Nuckolls72} J. Nuckolls \textit{et al.}, Nature (London) 239, 139 (1972).
 \bibitem{Atzeni04} S. Atzeni and J. Meyer-ter-Vehn, \textit{The Physics of Inertial Fusion: Beam plasma interaction, hydrodynamics, dense plasma physics} (Clarendon Press, Oxford, 2004).
 \bibitem{Bodner98} S. E. Bodner \textit{et al.}, Phys. Plasmas 5, 1901 (1998).
 \bibitem{Lindl95} J. Lindl \textit{et al.}, Phys. Plasmas 2, 3933 (1995).
 \bibitem{Taylor50} G. Taylor, Proc. R. Soc. London. Ser. A 201, 192 (1950).
 \bibitem{Haan11} S. W. Haan \textit{et al.}, Phys. Plasmas 18, 051001 (2011).
 \bibitem{Lindl14} J. Lindl \textit{et al.}, Phys. Plasmas 21, 020501 (2014).
 \bibitem{Hurricane14} O. Hurricane \textit{et al.}, Nature 13008 (2014).
 \bibitem{Dittrich14} T. R. Dittrich \textit{et al.}, Phys. Rev. Lett. 112, 055002 (2014).
 \bibitem{He13} X. T. He \textit{et al.}, Plenary talk at the IFSA2013, September 09-13, Nara, Japan (2013). Also, in Proceedings of IFSA (2013);
 \bibitem{Fan13} Z. F. Fan \textit{et al.}, arXiv:1303.1252[physics plasm-pj] (2013).
 \bibitem{Krokhin73} O. N. Krokhin and V.B.Rozanov, Sov. J. Quantum Electron 2, 393(1973).
 \bibitem{Olson11} R. Olson \textit{et al.}, Phys. Plasmas 18, 032706 (2011).
 \bibitem{Lan14} K. Lan \textit{et al.}, Phys. Plasmas 21, 010704 (2014);
 \bibitem{Lan14-2} K. Lan \textit{et al.}, Phys. Plasmas 21, 052704 (2014).
 \bibitem{Kline13} J. L. Kline \textit{et al.}, Phys. Plasmas 20, 056314 (2013).
 \bibitem{Town14} R. P. J. Town \textit{et al.}, Phys. Plasmas 21, 056313(2014).
 \bibitem{Temporal06} M. Temporal \textit{et al.}, Phys. Plasmas 13, 122701 (2006).
 \bibitem{Keskinen09} M. Keskinen, Phys. Rev. Lett. 103, 055001(2009).
 \bibitem{SGII14} In the experiment, the spacial nonuniformity induced by the deposition of lasers is better smoothed by a supersonic electron-ablatively heat wave, which is produced by intense lasers irradiating the critical-density foam.
 \bibitem{Doppner12} T. Doppner \textit{et al.}, Phys. Rev. Lett. 108, 135006 (2012).
 \bibitem{Wallace99} J. Wallace \textit{et al.}, Phys. Rev. Lett. 82, 3807 (1999).
 \bibitem{Betti07} R. Betti \textit{et al.}, Phys. Rev. Lett. 98, 155001 (2007).
 \bibitem{Theobald12} W. Theobald \textit{et al.}, Phys. Plasmas 19, 102706 (2012).


\end{thebibliography}
\end{document}